\documentstyle[twoside,fleqn,espcrc2]{article}

\def\hc{\hbox{h.c.}}

\newcommand{\AmS}{{\protect\the\textfont2
  A\kern-.1667em\lower.5ex\hbox{M}\kern-.125emS}}

\title{The twisted Wilson fermion for the Standard Model 
on a lattice}

\author{
She-Sheng Xue
\address{
Physics Department, University of Milan, Via Celoria 16, Milan, Italy}}
       
\begin{document}

\begin{abstract}
We present a preliminary study of a chiral-gauge-invariant Wilson term for the
Standard Model on a lattice. In the strong coupling limit for doublers, this
Wilson term shows that the doublers of leptons (baryons) are anti-baryons
(anti-leptons), and they are decoupled by getting gauge invariant masses. We
discuss that the possibility of achieving the low-energy Standard Model could
be realized by the ``three-fermion-cuts'' at the weak coupling limit for normal
modes, where baryons and anti-baryons dissolve into their constituents. 
 
\end{abstract}

\maketitle 

\section{Twisted pairing phenomenon in the SM}

The first generation of the SM contains 15 Weyl fermions. 
The left-handed sector comprises of 8 Weyl fermions:
\begin{eqnarray}
&&\nu_L,\hskip0.3cm e_L;\hskip0.5cm u_L,\hskip0.3cm d_L\label{l}\\
&&\nu^c_L,\hskip0.3cm e^c_L;\hskip0.5cm u_L^c,\hskip0.3cm d_L^c,
\label{al}
\end{eqnarray}
where the subscript ``$c$'' indicates 
anti-particles that are right-handed, and the color index is dropped. 
A right-handed neutrino $\nu_R$ is added into the right-handed sector,
$L\rightarrow R$ in eqs.(\ref{l},\ref{al}) and henceforth for all equations.

By using the 8 left-handed and 8 right-handed elementary Weyl particles and their
anti-particles, we can construct 8 right-handed and 8 left-handed 
Weyl three-fermion-states (3FS) called as ``proton'',
``neutron'' and leptoquarks. In the 
symmetries of the SM , these 8+8 3FS carry 
the exactly same quantum numbers as 8+8 elementary Weyl fermions,
however, their chiralities are just opposite. 

We take the ``proton'' and ``neutron'' as an example to show how these
3FS are constructed (color index is anti-symmetrized),
\begin{eqnarray}
P_L(u^c,u,d)&\sim& (\bar u_L^c\cdot u_L)d_L+\cdot\cdot\cdot ,\label{p}\\
N_L(d^c,d,u)&\sim& (\bar d_L^c\cdot d_L)u_L+\cdot\cdot\cdot .\label{n}
\end{eqnarray}
One can check that in the gauge symmetries of the SM, the following pairs:
\begin{eqnarray}
P^c_L \leftrightarrow e_L;&\hskip0.2cm&
N^c_L \leftrightarrow \nu_L,\label{nu}\\
P_L \leftrightarrow e^c_L;&\hskip0.2cm&
N_L \leftrightarrow \nu^c_L,\label{nu'}
\end{eqnarray}
have the same quantum numbers and opposite chiralities.

This is what we call the twisted pairing phenomenon in the SM. 
In fact, this twisted pairing phenomenon is the vector-like phenomenon 
required by the ``no-go'' theorem of Nielsen-Ninomiya\cite{nn}. This is just the 
consequence of the anomaly-free in the SM. 

Via the Eichten-Preskill mechanism\cite{ep}, these 3FS are 
bound states due to strong multi-fermion vertices that we give as\footnote{
The vertex that breaks $Z_3\otimes Z_3$ symmetry is given by M.~Creutz in the
same proceeding.}
\begin{equation}
V_L\!=\!\bar e_L\cdot P_L^c\!+\!\bar P_L\cdot e^c_L\!+\!
\bar \nu_L\cdot N_L^c\!+\!\bar N_L\cdot \nu^c_L\!+\!\hc, \label{lc}
\end{equation}
which preserve the gauge symmetries of the SM and violate
the $U_{B+L}(1)$ global symmetry. 

These vertices couple 16+16 Weyl
fermions for giving rise to 8+8 mass terms of 
Majorara-type. However, these mass terms are not only $SU_L(2)\otimes U_Y(1)$
invariant, but also $U_{em}(1)$ and $SU_c(3)$ invariant.
In the Domain-wall fermion\cite{k}, we study these vertices to show this
is indeed the case\cite{cx}.

\section{Twisted Wilson terms for doublers}

Going onto the lattice regularization of the SM, we have the 
doubling problem for each species of the 16 Weyl fermions (\ref{l}).
We introduce the twisted Wilson terms for doublers,
\begin{equation}
V_L\!=\!\bar e_L\Delta P_L^c\!+\!\bar P_L\Delta e^c_L\!+\!
\bar \nu_L\Delta N_L^c\!+\!\bar N_L\Delta \nu^c_L\!+\!\hc ,\label{wlc}
\end{equation}
where $\Delta$ is the ordinary second order derivatives in the lattice 
with appropriate gauge fields to preserve the $SU_L(2)\otimes U_Y(1)$ symmetries.
Analogous to the Wilson term, we expect that
the twisted Wilson terms give different gauge-invariant masses to different
doublers.

In the strong coupling 
limit $g\gg 1$, we can calculate the two-point functions for doublers
($p\simeq \pi_A$). As an example, We show the result for the sector of electron 
and anti-proton, 
\begin{eqnarray}
\int_x\langle e_L(0),\bar e_L(x)\rangle\!&=\!&{{1\over a}\gamma_\mu\sin p_\mu
P_R\over{1\over a^2}\sin^2p_\mu+g^2w^2(p)},\label{p1}\\
\int_x\langle P^c_L(0),\bar P^c_L(x)\rangle &\!=&\!
{{1\over a}\gamma_\mu\sin p_\mu
P_L\over{1\over a^2}\sin^2p_\mu+g^2w^2(p)},\label{p2}\\
\int_x\langle e_L(0),\bar P^c_L(x)\rangle \!&=\!&{
gw(p)P_L\over{1\over a^2}\sin^2p_\mu+g^2w^2(p)},\label{p3}
\end{eqnarray}
where $w(p)=\sum_\mu(1-\cos p_\mu)$ and $\int_x=\int_xe^{-ipx}$. 

The anti-proton ($P^c_L$) 
indeed behaves as a simple pole ($p\simeq\pi_A$) in its propagator (\ref{p2}). 
Eq.(\ref{p3}) 
shows that the electron ($e_L$) and anti-proton ($P_L^c$) mix up to 
form a massive ``Dirac'' fermion $\psi_{e_L}(x)=e_L(x)+P^c_L(x)$,
and its propagator is ($p\simeq\pi_A$),
\begin{equation}
\int_x\langle \psi_{e_L}(0),\bar \psi_{e_L}(x)\rangle =
{{1\over a}\gamma_\mu\sin p_\mu+gw(p)
\over{1\over a^2}\sin^2p_\mu+g^2w^2(p)}.\label{dp}
\end{equation}
The doublers are thus decoupled.

In the basis of the twisted Wilson terms and the spectrum for doublers 
($p\simeq\pi_A$) (\ref{dp}), we find that
the doublers of the electron ($e_L$) are the anti-proton ($P^c_L$) at 
$p\simeq\pi_A$. Actually, the doublers of the electron ($e_L$) are 
the anti-quark $d_L^c$ at $p\simeq\pi_A$,
since the anti-proton is given by
\begin{eqnarray}
P^c_L(p)&\!\sim&\! \int_{p_1,p_2}\left[\bar u_L(p_1)\!\cdot\! u_L^c(p_2)\right]
d_L^c(p_3)\!+\!
\cdot\cdot\cdot;\label{pp}\\
p_3\!&=&\!p+p_2-p_1\sim p\simeq\pi_A,\label{m}
\end{eqnarray} 
which is composed by the anti-quark $d^c_L$ at $p\simeq\pi_A$ and 
a soft ($p_1-p_2\sim 0$) pair of the quark $\bar u_L(p_1)$ and 
anti-quark $u^c_L(p_2)$, where the $p_1$ and $p_2$ must be in the same 
Brillouin zone. 

On the other hand, 
the doublers of the quark $d_L$ are the
leptoquark [$(\bar u_L\cdot u^c_L)e^c_L$] at $p\simeq\pi_A$, and this
means that the doublers of the quark $d_L$ are the anti-electron ($e_L^c$)  
at $p\simeq\pi_A$. 
Similar discussions can be applied to the neutrino-uquark
sector. In the Brillouin zones ($p\simeq\pi_A$),
we have,
\begin{eqnarray}
e_{L,R}\rightarrow d^c_{L,R}&\hskip0.3cm &d_{L,R}\rightarrow e^c_{L,R}
\label{d1}\\
\nu_{L,R}\rightarrow u^c_{L,R}&\hskip0.3cm &u_{L,R}\rightarrow \nu^c_{L,R}.
\label{d2}
\end{eqnarray}
In the twisted Wilson terms, the doublers of leptons and quarks (together with
an appropriate pair of quark and anti-quark) are twisted to 
preserve the 
chiral gauge symmetries of the SM. This is in agreement with the ``no-go''
theorem.

\section{The shift-symmetries of $e_{L,R}$ and $\nu_{L,R}$}

The vertices (\ref{wlc}) possess the shift symmetries\cite{gp} of $e_{L,R}$ and 
$\nu_{L,R}$. The Ward identities corresponding to these 
symmetries are, e.g.,
\begin{eqnarray}
{i\over2a}\gamma_\mu\partial^\mu e'_L(x)+g\Delta P'^c_L(x)
-{\delta\Gamma\over\delta\bar e'_L(x)}\!&=\!&0;\label{w1}\\
{i\over2a}\gamma_\mu\partial^\mu \nu'_L(x)+g\Delta N'^c_L(x)
-{\delta\Gamma\over\delta\bar \nu'_L(x)}\!&=\!&0,\label{w2}
\end{eqnarray}
where $\Gamma$ is the vacuum functional 
and
$\langle\cdot\cdot\cdot\rangle$ is the vacuum expectation value.
The primed fields $\psi'=\langle\psi\rangle$.
The Ward identities (\ref{w1},\ref{w2}) are helpful to find 1PI vertices that
describe the dynamics we are looking for in the phase diagram.

Based on these Ward identities, we can obtain the
various identities,
\begin{eqnarray}
\int_x{\delta^2\Gamma\over\delta e'_L(x)\delta\bar e'_L(0)}&=&
{i\over a}\gamma_\mu\sin p^\mu P_L;\label{w3}\\
\int_x{\delta^2\Gamma\over\delta \nu'_L(x)\delta\bar \nu'_L(0)}&=&
{i\over a}\gamma_\mu\sin p^\mu P_L,\label{w4}
\end{eqnarray}
which show that lepton fields do not receive the wave-function 
renormalization $Z_3$, and
\begin{eqnarray}
\int_x{\delta^2\Gamma\over\delta e'_R(x)\delta\bar e'_L(0)}&=&{1\over2}
\Sigma_e(p)=0,\label{w6}\\
\int_x{\delta^2\Gamma\over\delta \nu'_R(x)\delta\bar \nu'_L(0)}&=&
{1\over2}\Sigma_\nu(p)=0.\label{w7}
\end{eqnarray}
These equations are independent of the coupling $g$. Analogously, we can also 
obtain the vacuum expectation values $\langle \bar q_R\cdot l_L\rangle =0$, 
where $q=u,d$ and $l=\nu,e$, and $L\rightarrow R;R\rightarrow L$.

\section{Hard spontaneous symmetry breaking ?}

As for the vacuum expectation values $\langle
\bar q_L\cdot q'_R\rangle$, where $q=u,d$ and $q'=u,d$, 
we expect that they are zero, since they are the vacuum 
expectation values
of connections between the left-vertex $V_L$ and the 
right vertex $V_R$, 
\begin{equation}
\langle\bar q_L\cdot q'_R\rangle \sim\langle\bar l_L\cdot l'_R\rangle=0,
\label{lr1}
\end{equation}
where $l,l'=\nu,e$. This is due to
the fact that the left-handed fields and the right-handed
fields are completely separated and the shift-symmetry protects 
such separation. 

However, for the weak multi-fermion coupling ``$g$'', a spontaneous symmetry 
breaking is expected. The vacuum expectation values of fields within the 
L-(or R-) sector, e.g. $\langle\bar q_L\cdot q^c_L\rangle,\cdot\cdot\cdot$ 
break symmetries. We need to do dynamical calculations to find where the broken 
phase is.

We expect a symmetric segment ($g_c<g<\infty$)\cite{xue97} in the strong 
coupling ``$g$'' phase, where normal
fermion modes could get around the broken phase and not form the 
3FS. Analytical and numerical 
work is needed to verify that such a symmetric ``segment'' indeed exists.

\section{Chiral fermions at three-fermion-cuts}

The most important question is what
the spectrum for normal modes ($p\sim 0$) in the continuum limit would be in 
this symmetric segment. There are two possible cases.

\begin{enumerate}
\begin{itemize}
\item the spectrum is vector-like;
\item at the scale $\epsilon$ ($250$GeV$<\epsilon\ll{\pi\over a}$) for the 
continuum limit, the 3FS dissolve into the three-fermion-cuts 
that are the virtual states of three free chiral fermions and have the same 
gauge quantum numbers of the 3FS.\cite{xue}.
\end{itemize}
\end{enumerate}

We are probably ended up with the first case,
in which 3FS are simple poles for $p\sim 0$.
A plausible study\cite{ys} for this case to happen bases on the locality of the 
theory that leads to the continuation of
the simple poles, i.e.~3FS, in the whole Brillouin zone. 

Nevertheless, we try to argue for the second case. In the continuum limit
$p\rightarrow 0$, the effective multi-fermion coupling $g(p)$ goes to zero, the
3FS are no longer simple poles, but dissolve into three-fermion-cuts on the
physical sheet, where the ``no-go'' theorem does not apply. This is indeed
physically feasible. 

Let us look at the wave-function renormalization $Z_3$'s of 
the 3FS, which characterize the 3FS propagating
as particles (simple poles), e.g.,
\begin{equation}
\int_x\langle P^c_L(0), \bar P^c_L(x)\rangle\sim {Z^2_3\over 
\slash p + m(p)}P_R.
\label{z3}
\end{equation} 
$Z_3$'s can be calculated by the Ward identities (\ref{w1},\ref{w2}),
\begin{eqnarray}
Z_3^p(p)=\int_x{\delta^2\Gamma\over\delta P'^c_L(x)\delta\bar e'_L(0)}\!&=&\!
gw(p),\label{w8}\\
Z_3^n(p)=\int_x{\delta^2\Gamma\over\delta N'^c_L(x)\delta\bar \nu'_L(0)}\!&=&\!
gw(p).\label{w9}
\end{eqnarray}
And $
Z_3^p(p)\rightarrow
0,\hskip0.2cm
Z_3^n(p)\rightarrow
0$ for $p\rightarrow 0$. This indicates that the 3FS 
at $p\sim 0$ increase their size, and 
would turn into cuts.

\end{document}